\RequirePackage{fix-cm}
\documentclass[smallextended]{svjour3}       % onecolumn (second format)
\smartqed  % flush right qed marks, e.g. at end of proof
\usepackage{graphicx}
\begin{document}

\title{Energy dependence of $\bar{K}N$ interaction in nuclear medium%\thanks{Grants or other notes
}


\author{Ale\v{s} Ciepl\'{y}
}


\institute{A. Ciepl\'{y} \at
              Nuclear Physics Institute, 250 68 \v{R}e\v{z}, Czech republic \\
              Tel.: +420-266173284\\
              Fax: +420-220940165\\
              \email{cieply@ujf.cas.cz}           %  \\
}

\date{Received: date / Accepted: date}

\maketitle

\begin{abstract}
When the $\bar{K}N$ system is submerged in nuclear medium the $\bar{K}N$ 
scattering amplitude and the final state branching ratios exhibit 
a strong energy dependence when going to energies below the $\bar{K}N$ 
threshold. A sharp increase of $\bar{K}N$ attraction below the $\bar{K}N$ 
threshold provides a link between shallow $\bar{K}$-nuclear potentials 
based on the chiral $\bar{K}N$ amplitude evaluated at threshold 
and the deep phenomenological optical potentials obtained in fits 
to kaonic atoms data. We show the energy dependence of the in-medium 
$K^{-}p$ amplitude and demonstrate the impact of energy dependent 
branching ratios on the $\Lambda$-hypernuclear production rates.
\keywords{kaon-nucleon amplitude \and nuclear medium \and hypernuclei}
\end{abstract}

\vspace*{5mm}
A key issue in studying in-medium $K^-$ meson interactions concerns 
the strength of the attractive $K^-$ nuclear potential.  
Chirally based coupled channels calculations lead to $K^-$ nuclear optical 
potentials that are about 80 MeV deep at nuclear densities 
$\rho = \rho_0 = 0.17~{\rm fm}^{-3}$. The incorporation of kaon selfenergy 
reduces the depth of the potential to approximately $40-50$ MeV while much
deeper potentials, in the range ${\rm Re}\:V_{K^-} (\rho_0)\sim -$(150--200) MeV 
are obtained in comprehensive global fits to kaonic atoms data \cite{1994FGB}. 
In our recent paper \cite{2011CFGGM} we reported on a new self consistent 
treatment of chirally motivated $K^-N$ amplitudes that lead to deep 
$K^-$ nuclear potentials, considerably deeper than the `shallow' 
potentials deduced in earlier models based on the chiral amplitude 
\cite{1996WKW}, \cite{2000RO}. 

The elementary $\bar{K}N$ interaction at threshold and low energies 
is well understood within chiral models combined with the coupled channels 
re-summation techniques. In our approach we employ chirally motivated 
coupled-channel potentials that are taken in a separable form \cite{2010CS}.
The meson-baryon channels are composed from the $\pi\Lambda$, $\pi\Sigma$, 
$\bar{K}N$, $\eta\Lambda$, $\eta\Sigma$, and $K\Xi$ states (taken with all 
appropriate charge combinations). The parameters of the model, the inverse 
ranges that define the off-shell Yamaguchi form factors and low energy constants  
of the SU(3) chiral Lagrangian that contribute to the inter-channel couplings, 
are fitted to the kaonic hydrogen and low energy $K^{-}p$ reactions data.
A detailed description of our model can be found in Refs.~\cite{2010CS} 
and \cite{2011CFGGM}.
 
\begin{figure}
\centering 
\includegraphics[width=0.95\textwidth]{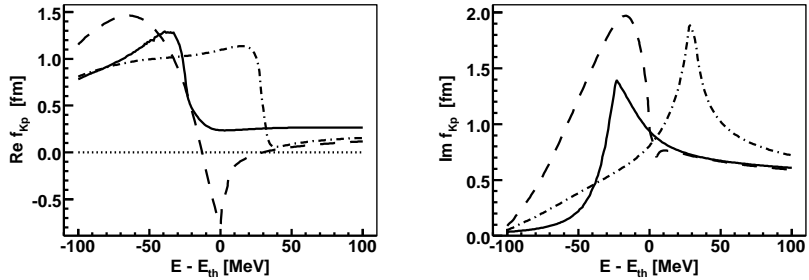}
\caption{
Energy dependence of the $K^{-}p$ scattering amplitude. The left and 
right panels refer to the real and imaginary parts of the amplitude, 
respectively, and $E_{th}$ denotes the $K^{-}p$ threshold energy. 
Dashed curves: free space, dot-dashed: Pauli blocking in nuclear medium 
at $\rho = \rho_0$, solid curves: combined effect of Pauli blocking 
and hadron selfenergies at $\rho = \rho_0$.
}
\label{fig:aKp}       % Give a unique label
\end{figure}

When the $\bar{K}N$ system is submerged in nuclear medium the $\bar{K}N$ 
scattering amplitude and the final state branching ratios exhibit 
a strong energy dependence when going to energies below the $\bar{K}N$ 
threshold. In Fig.~\ref{fig:aKp} we show an energy dependence of the elastic 
scattering amplitudes $f_{K^-p}$ in free space and for two versions 
of in-medium modifications, performed with a leading order chiral 
model labeled as TW1 in Ref.~\cite{2011CFGGM}. The free space
$K^-p$ amplitude exhibits a typical structure related to an $I=0$ $\bar{K}N$ 
quasi-bound state which is assigned to a well known $\Lambda(1405)$ resonance.
When the $K^-p$ system is submerged in nuclear medium the Pauli blocking 
shifts the structure to higher energies, about $30-40$ MeV 
above the $\bar{K}N$ threshold. However, the addition of kaon selfenergies 
(we also included baryon and pion selfenergies in our model, though their 
impact is minute) brings it back below the threshold as demonstrated by 
the solid curves in the figure. In this regime the most striking 
feature is a sharp increase in the real part of the amplitude when going 
to subthreshold energies. Consequently, the $K^{-}p$ interaction becomes 
much stronger at energies about 30 MeV below the $K^{-}p$ threshold with 
respect to its strength at threshold. We have shown \cite{2011CFGGM} that 
this is exactly the region of energies probed by kaons at the lowest 
$K$-atomic orbits, so the $K^{-}$-nuclear optical potential becomes 
much deeper than when it were constructed from the amplitudes taken at 
the $\bar{K}N$ threshold. 

\begin{figure}
\centering 
\includegraphics[width=0.95\textwidth]{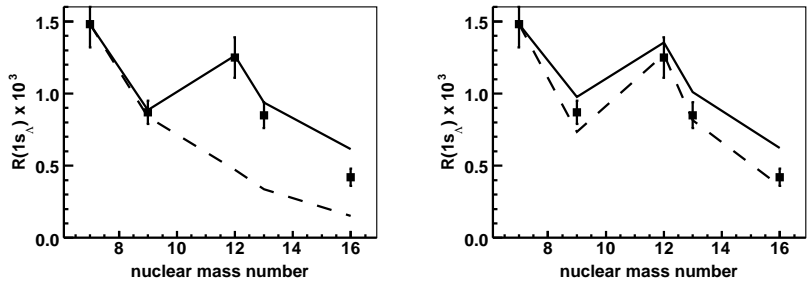}
\caption{The $A$-dependence of the $1s_{\Lambda}$ hypernuclear formation rates, 
experimental data from Ref.~\cite{2011FINUDA}. The theoretical rates are normalized 
to the $^{7}$Li experimental value and were calculated with a phenomenological density 
dependent kaon-nuclear optical potential (dashed lines) and with the chirally motivated 
$K^{-}$-nuclear optical potential (solid lines). Left panel: elementary 
BR($K^-N\rightarrow\pi\Lambda$) fixed at the threshold value for nuclear density 
$\rho = \rho_{0}/2$, right panel: energy and density dependent BR.}
\label{fig:rates}       
\end{figure}

The impact of the in-medium $\bar{K}N$ dynamics and energy dependence 
of the $\bar{K}N$ amplitudes on the characteristics of kaonic atoms and 
kaon-nuclear states was investigated in Ref.~\cite{2011CFGGM}. 
Here we mention another application to the $\Lambda$-hypernuclear production 
in the ($K^-_{\rm stop}, \pi^{-}$) reactions. Recently, the production rates were 
established by the FINUDA collaboration \cite{2011FINUDA} for five $p$-shell nuclear 
targets from $^{7}$Li to $^{16}$O. Since the absolute normalization of the experimental capture rates
is a delicate matter and the calculated rates are generally much 
lower than the measured ones, we have focused on the $A$-dependence
of the $1s_{\Lambda}$ formation rates \cite{2011CFGK}. 
The Figure \ref{fig:rates} shows the effect of energy 
dependent branching ratios BR($K^{-}n \rightarrow \pi^{-}\Lambda$) on the 
rates computed within the framework of the distorted wave impulse 
approximation. There, the $K^{-}$-nuclear optical potential is used to 
generate the $K$-atomic initial state wave function. 
In the left panel (which was not published earlier) of Fig.~\ref{fig:rates} 
we present the rates calculated with the BR fixed at a value obtained 
at the $\bar{K}N$ threshold while the theoretical rates in the right panel 
(taken from Ref.~\cite{2011CFGK}) were calculated with energy 
dependent BR averaged over the region of relevant subthreshold energies, 
specific for each nuclear target. With the BR fixed at the $\bar{K}N$ 
threshold the $A$-dependence is reproduced much better by the chirally 
motivated optical potential which is shallow at the $\bar{K}N$ threshold. 
Since both the phenomenological and chirally motivated optical potentials 
are sufficiently deep at the subthreshold energies relevant for the evaluation 
of the energy dependent BR, they lead to similar $1s_{\Lambda}$ formation rates.

\vspace*{-2mm}
\begin{acknowledgements}
The author acknowledges a fruitful collaboration with E. Friedman, A. Gal, D. Gazda, J. Mare\v{s}, 
V. Krej\v{c}i\v{r}\'{\i}k and J. Smejkal who coauthored the papers the report is based on.
This work was supported by the GACR Grant No. 202/09/1441.
\end{acknowledgements}
\vspace*{-2mm}



\end{document}